\documentclass[journal,twoside,web]{ieeecolor}
\usepackage{generic}
\usepackage{cite}
\usepackage{amsmath}
\usepackage{array}
\usepackage{graphicx}
\usepackage{bm}
\usepackage{float}
\usepackage[hyphens,spaces,obeyspaces]{url}
\usepackage{makecell}
\usepackage{adjustbox}
\usepackage{booktabs}
\usepackage{dcolumn}
\usepackage{upgreek}
\usepackage{dblfloatfix}
\begin{document}

\bstctlcite{IEEEexample:BSTcontrol}

\title{Analysis of Ferroelectric Negative Capacitance-Hybrid MEMS Actuator Using Energy-Displacement Landscape}

\author{Raghuram Tattamangalam Raman, Jeffin Shibu, Revathy Padmanabhan and Arvind Ajoy   \thanks{R. Tattamangalam Raman, J. Shibu, R. Padmanabhan and A. Ajoy are with Electrical Engineering, Indian Institute  of Technology
Palakkad,             Palakkad,             India.             e-mail:
121704004@smail.iitpkd.ac.in; 121801055@smail.iitpkd.ac.in; revathyp@iitpkd.ac.in; arvindajoy@iitpkd.ac.in}
}

\maketitle

\begin{abstract}
We propose an energy-based framework to analyze the statics and dynamics of a ferroelectric negative capacitance-hybrid Microelectromechanical System (MEMS) actuator. A mapping function that relates the  charge on the ferroelectric to displacement of the movable electrode, is used to obtain the Hamiltonian of the hybrid actuator in terms of displacement. We then use graphical energy-displacement and phase portrait plots to analyze static pull-in, dynamic pull-in and pull-out phenomena of the hybrid actuator.  Using these, we illustrate the low-voltage operation of the hybrid actuator to static and step inputs, as compared to the standalone MEMS actuator. The results obtained are in agreement with the analytical predictions and numerical simulations. The proposed framework enables straightforward inclusion of adhesion between the contacting surfaces, modeled using van der Waals force. We show that the pull-in voltage is not affected, while the pull-out voltage is reduced due to adhesion. The proposed framework provides a physics-based tool to design and analyze  negative capacitance based low-voltage MEMS actuators.\end{abstract}

\begin{IEEEkeywords}
Microelectromechanical System (MEMS), Electrostatic MEMS Actuator, Ferroelectric Negative Capacitance, Energy-Displacement Landscape, Phase-portrait.
\end{IEEEkeywords}

\section{Introduction}
\label{sec:intro}
\IEEEPARstart{E}{lectrostatic} MEMS (Microelectromechanical System) actuators are of great interest in modern electronic applications like Radio Frequency (RF) MEMS switches and	digital micromirror devices \cite{choudhary2016mems,gad2005mems,rebeiz2004rf}. While these devices are energy efficient,  they demand high operating voltages \cite{rebeiz2004rf}. Ultra-scaled MEMS devices can operate at low-voltages, but are challenging to fabricate reliably due to effects such as stiction \cite{pawashe2013scaling}. Masuduzzaman and Alam \cite{masuduzzaman2014effective} proposed a novel approach to reduce the operating voltage of a MEMS device without scaling its air-gap. Their $\emph{hybrid}$ MEMS actuator consists of  a ferroelectric capacitor, exhibiting negative capacitance, connected in series with the MEMS actuator. This idea  is similar  to the development of  Negative Capacitance-Field Effect Transistors (NC-FETs), proposed originally in Ref. \cite{salahuddin2008use}. Readers are directed to  Refs. \cite{kobayashi2018perspective, tu2018ferroelectric} for a review of NC-FETs. Low-voltage operation is predicted to arise due to  the internal voltage amplification when a negative capacitance is connected in series with a positive capacitance.  Different signatures of  negative capacitance in ferroelectrics have been experimentally reported -- for example, a charge-voltage curve  with negative slope \cite{khan2015negative, HoffmannNature2019}, an enhanced total capacitance \cite{asif2011experimental}, and  steady state charge boost \cite{tasneem2021differential} in a ferroelectric-dielectric heterostructure.

No experimental realizations of the hybrid actuator have been reported so far. Many authors have analyzed the performance of these devices through analytical and numerical techniques. The ferroelectric capacitor is governed by the nonlinear Landau-Khalatnikov equation \cite{masuduzzaman2014effective}, which relates the voltage across the ferroelectric to its charge. The MEMS actuator, on the other hand, is governed by a nonlinear  differential equation \cite{younis2011mems}, expressed in terms of displacement of the movable electrode. It is convenient to describe both the ferroelectric and the MEMS actuator in terms of a common entity. For instance, the response  of the hybrid actuator to slowly varying (quasi-static) inputs  was analytically studied in Refs. \cite{masuduzzaman2014effective,choe2017adjusting} using charge as the common variable. They solve the algebraic equations that describe the balance between the  electrostatic attraction and spring restoring  forces at equilibrium. Our earlier work \cite{raghu2020charge} analyzes both the static and dynamic response  of standalone MEMS actuators based on their energy-charge landscape. This technique can, in principle,  be extended to analyze the hybrid actuator as well. 

However,  for many applications, displacement is a more natural coordinate used to analyze MEMS actuators \cite{younis2011mems}. Analysis based on displacement is convenient to include effects such as adhesion \cite{decuzzi2006bouncing,granaldi2006dynamic,chakraborty2011experimental} and a non-linear spring  \cite{nemirovsky2001methodology}, that are directly described in terms of displacement. We had developed \cite{raghu2020spice} a numerical model to analyze both the statics and dynamics of the hybrid actuator, based on displacement. The numerical model solves the  nonlinear, coupled,  differential equations using the inbuilt solvers of a circuit simulator. The numerical approach, though, provides very limited physical insight into the response of the hybrid actuator.

In this work, we develop a physics-based graphical framework, using displacement as the dynamical variable, that facilitates a systematic analysis of the statics and dynamics of the hybrid actuator.  We employ a coordinate transformation from the charge to the displacement of the movable electrode, in order to describe the ferroelectric in terms of displacement. This allows us to express the Hamiltonian (energy) of the hybrid MEMS actuator in terms of displacement. We then  use graphical energy-displacement and phase portrait (velocity vs. displacement) plots to investigate static pull-in, dynamic pull-in and pull-out phenomena of the hybrid MEMS actuator.  

The usefulness of describing the Hamiltonian in terms of displacement is illustrated by studying the effect of adhesion in the hybrid actuator. Adhesion plays a major role when the top electrode comes in contact with the bottom surface. We include the effect of adhesion between contacting surfaces, by adding a term corresponding to the van der Waals force \cite{decuzzi2006bouncing,granaldi2006dynamic,chakraborty2011experimental} into the Hamiltonian of the system. We show that adhesion reduces the pull-out voltage but does not affect the pull-in voltage. We demonstrate how the actuator can be redesigned so that the reduction in pull-out voltage (due to adhesion) can be compensated. We show that this redesign causes an increase in the pull-in voltage; nevertheless, the new pull-in voltage is predicted to still be considerably lower than the pull-in voltage of the standalone MEMS actuator. Since the proposed energy framework uses only graphical plots for the analysis, this serves as a quick design and analysis tool to predict the pull-in and pull-out behaviour of the hybrid actuator.



\begin{figure}[t]
    \centering
    \includegraphics[scale=0.8]{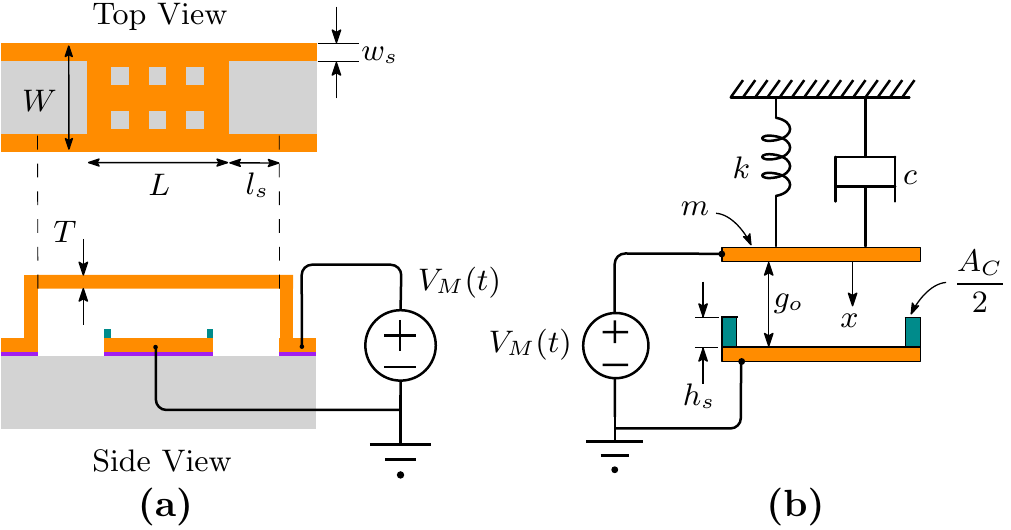}
    \caption{(a) Schematic representation of the standalone clamped-clamped
MEMS actuator. (b) Equivalent 1-DOF model of the MEMS actuator.}
    \label{fig:MEMS_1DOF}
\end{figure}

This article is organized as follows. Section \ref{sec:review} reviews the operation of the standalone and hybrid MEMS actuators.  Section \ref{sec:formalism} presents the Hamiltonian  of the hybrid actuator using the displacement  of the movable electrode as the dynamical variable. Section \ref{sec:design} discusses our design of the hybrid actuator. The static pull-in, dynamic pull-in and pull-out of the hybrid actuator are analyzed in Section \ref{sec:analysis}. Section \ref{sec:adhesion} investigates the effect of adhesion. Finally, Section \ref{sec:conclusion} presents our conclusions. 

\begin{table}[t]
\caption{\label{tab:Table1}%
Parameters of the hybrid MEMS actuator used in this work
}
	\centering
	\begin{tabular}{ll}
	\toprule
	\textbf{Parameter}&
	\textbf{Value}\\
	\midrule
	Beam material & Gold (Au) \cite{shekhar2017surface}\\
	Length of the beam, $L$ & 140 $\mu$m \\
	Width of the beam, $W$ & 120 $\mu$m\\
	Actuation area, $A_M$ & 1.44 $\times$ $10^{-8}$~m$^2$\\	
	Young's modulus, $E$ & $78$ GPa\\
	Density, $D$ & $19280$ kg/m$^3$~ \\
	Mass, $m = 0.35 \times D \times$ volume & $5.6 \times 10^{-11}$  kg \cite{rebeiz2004rf}\\
	Width of the support, $w_S$ & 20 $\mu$m\\
	Length of the support, $l_S$ & 80 $\mu$m\\
	Thickness, $T$ & $0.5~\mu$m\\
	Spring constant, $k = 4 E w_s(\frac{T}{l_s})^3$ & $1.52$ N/m \cite{rebeiz2004rf}\\
	Initial air-gap, $g_o$ & $2~\mu$m\\
	Stopper height, $h_s$ & $0.15~\mu$m\\
	Area of contact, $A_C$ & $16~\mu$m$^2$ \cite{granaldi2006dynamic} \\
	Permittivity of free space, $\epsilon_o$ & $8.854 \times 10^{-12}$ F/m \\
	\midrule
	Ferroelectric material & HfO$_2$ \cite{kobayashi2016device}\\
	$\alpha_F$ & $-2.88 \times 10^{9}$ m/F\\
	$\beta_F$ & $3.56 \times 10^{11}$ m$^5$/F/C$^2$\\
	$\gamma_F$ & $0$ m$^9$/F/C$^4$\\
	Ferroelectric thickness, $t_F$ & $45.24$ nm\\
	Ferroelectric area, $A_F$ & $9.87 ~\mu$m$^2$\\
	\bottomrule
	\end{tabular}
\end{table}
\section{Review of Standalone and Hybrid MEMS Actuators} \label{sec:review}	
We consider a standalone electrostatic MEMS actuator as a clamped-clamped beam with fixed-fixed flexure, as shown in Fig. \ref{fig:MEMS_1DOF}(a). Based on Refs. \cite{rebeiz2004rf,shekhar2017surface}, the values of the MEMS actuator parameters are listed in Table \ref{tab:Table1} and are fairly typical for MEMS actuators. These references  also detail the fabrication steps used to realize such MEMS actuators. The actuator is modeled using a single degree of freedom (1-DOF), parallel plate arrangement consisting of a pair of electrodes separated by an air-gap $g_o$, as shown in Fig. \ref{fig:MEMS_1DOF}(b). It is excited by an input voltage $V_M(t)$, where $t$ denotes time. The top electrode is movable and the bottom electrode is fixed. The stiffness of the beam, the inertia elements and the damping mechanisms of the actuator are effectively represented by a spring-mass-damper with spring constant $k$, mass $m$, and damping coefficient $c$. We assume that the displacement of the top electrode, denoted by a dynamical variable $x$, is limited by a pair of stoppers, as shown in Fig. \ref{fig:MEMS_1DOF}(b). These stoppers (with height $h_s$) are made of an insulating material and hence, prevent electrical short between the top and bottom electrodes \cite{iannacci2009measurement,giacomzzi2013rf}. The top electrode comes in contact with the stopper over an area $A_C$. We neglect damping in our analysis. The effect of surface forces is  neglected initially and included later in Section \ref{sec:adhesion}. 

\begin{table}[b]
\caption{\label{tab:Table2}%
Pull-in and pull-out of a standalone MEMS actuator. Values correspond to parameters listed in Table \ref{tab:Table1}. 
}
	\centering
	\begin{adjustbox}{width=\columnwidth,center}
	\begin{tabular}{lll}
	\toprule
	\textbf{Parameter}&
	\textbf{Expression \cite{younis2011mems}}&
	\textbf{Value}\\
	\midrule
	\vspace{3pt}
	Static pull-in voltage, $V_{SPI}$ & $\sqrt{8 k g_o^3/27 \epsilon_o A_M}$&$5.32$ V \\
	\vspace{3pt}
	Travel range, $X_{SPI}$ & $g_o/3$&0.67$~\mu$m\\
	\vspace{3pt}
	Dynamic pull-in voltage, $V_{DPI}$ & $\sqrt{k g_o^3/4 \epsilon_o A_M}$&$4.88$ V \\		\vspace{3pt}
	Dynamic pull-in displacement, $X_{DPI}$ & $g_o/2$&1$~\mu$m \\
	Pull-out voltage, $V_{PO}$ & $\sqrt{\frac{2~k~h_s^2~(g_o - h_s)}{\epsilon_o~A_M}}$&$1$ V \\
	System rise time, $t_{sys}$& $0.35 \times 2\pi \sqrt{m/k}$ & $13.28 ~\mu$s \\
	\bottomrule
	\end{tabular}
	\end{adjustbox}
\end{table}

The transient response of the MEMS actuator depends on the nature of the input voltage. The input voltage is considered to be \emph{slow} if its  rise time, $t_{inp}$, is significantly greater than the system rise time, $t_{sys}$ of the MEMS actuator. Empirically, $t_{sys}=0.35/f_0$ , where the resonant frequency $f_0 = \frac{1}{2 \pi} \sqrt{k/m}$ \cite{oberhammer2007accuracy}. Thus, when  $t_{inp} \gg  t_{sys}$, the actuator remains in quasi-static equilibrium. However, beyond a certain voltage, called the static pull-in voltage $V_{SPI}$, the movable electrode snaps down on to the fixed electrode. This condition is called static pull-in \cite{younis2011mems}. Consequently, the maximum distance travelled by the movable electrode, before it snaps down, is called the travel range $X_{SPI}$.  On the other hand, for a step input (with $t_{act}\ll t_{sys}$), the actuator is driven away from equilibrium. In the absence of damping, the response of the actuator is oscillatory, for voltages less than $V_{DPI}$, called the dynamic pull-in voltage. The maximum value of this oscillatory displacement of the electrode is referred to as the dynamic pull-in displacement, $X_{DPI}$. For any applied step voltage, greater than $V_{DPI}$, the movable top electrode snaps down onto the bottom electrode. This condition is called dynamic pull-in \cite{younis2011mems}. After achieving pull-in (static or dynamic), as the input voltage is reduced to a specific value, called the pull-out voltage $V_{PO}$, the pull-in condition is lost and the movable top electrode gets detached from the fixed bottom electrode. This condition is called pull-out \cite{younis2011mems}. Table \ref{tab:Table2} summarizes the expressions for pull-in/pull-out voltages and displacements for a standalone electrostatic MEMS beam, with zero damping. The corresponding values for the actuator described in Table \ref{tab:Table1} are also listed. Note that damping does not affect pull-out and static pull-in. However, with an increase in damping constant $c$, the dynamic pull-in voltage increases from $V_{DPI}$ and approaches the static pull-in voltage $V_{SPI}$ \cite{nielson2006dynamic}. Hence the analysis presented in this work (with $c=0$) provides an estimate of the lowest possible dynamical pull-in voltage. 

\begin{figure}[t]
    \centering
    \includegraphics[scale=1.0]{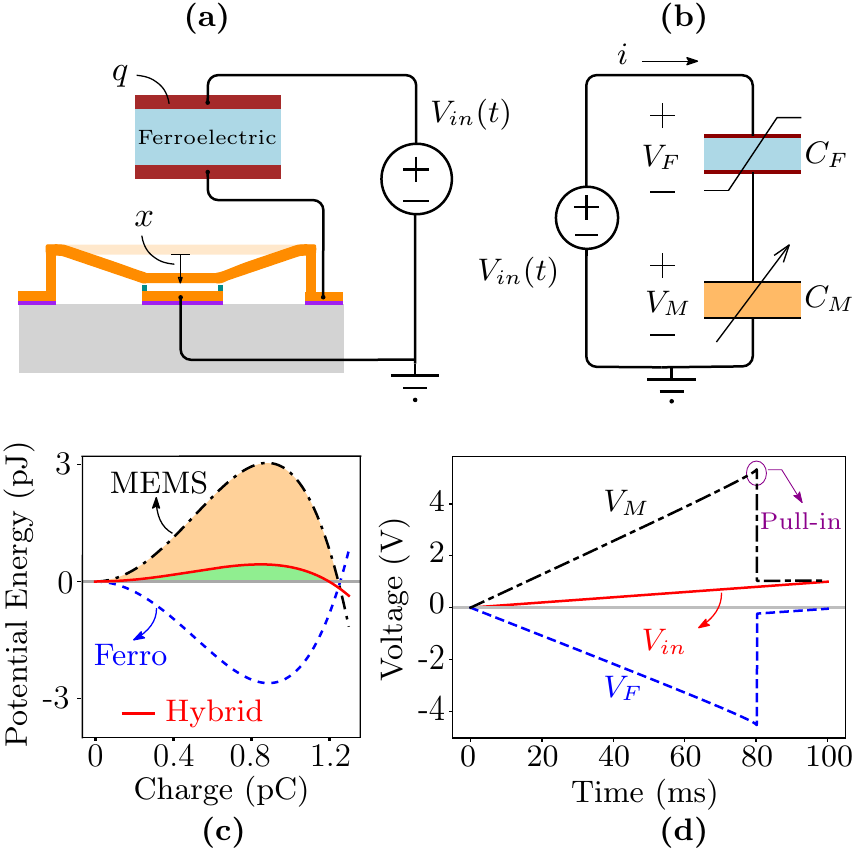}
    \caption{(a) Hybrid actuator formed by series connection of ferroelectric capacitor $C_F$ with the MEMS actuator. (b) Equivalent circuit representation. $C_M$ represents the variable capacitance provided by the MEMS actuator. (c) Potential energy-charge landscape plotted for $V_{in}=0$ V, depicting a lowered energy barrier (shaded green) in the hybrid actuator as compared to the MEMS actuator (shaded orange). (d) Voltages plotted as a function of time. The hybrid actuator operates at a lower voltage due to the voltage amplification ($V_M > V_{in}$) caused by the ferroelectric negative capacitance.}
    \label{fig:Hybrid_MEMS_Equiv}
\end{figure}

The hybrid actuator is formed by connecting  a ferroelectric capacitor $C_F$, in series with the standalone MEMS actuator, as shown in Fig. \ref{fig:Hybrid_MEMS_Equiv}(a). This configuration is similar to various negative capacitance transistors reported in literature, where the ferroelectric capacitor is connected externally to the transistor \cite{saeidi2020nanowire,saeidi2018effect,khan2017differential,khan2015negativeepi,choi2019negative,ko2017negative,jo2016negative}. The equivalent circuit representation of the hybrid actuator is shown in Fig. \ref{fig:Hybrid_MEMS_Equiv}(b), wherein the MEMS actuator is depicted as a variable capacitor $C_M$. The low voltage operation of the hybrid MEMS actuator can be understood from the potential energy-charge landscape (plotted for $V_{in}=0$ V in Fig. \ref{fig:Hybrid_MEMS_Equiv}(c)), following Ref. \cite{masuduzzaman2014effective}. The parameters assumed for the ferroelectric are listed in Table \ref{tab:Table1} and will be discussed later in Section \ref{sec:design}. Note that the energy barrier is lowered in the hybrid actuator  as compared to that of the standalone MEMS actuator. This observation is substantiated by numerical results shown in Fig. \ref{fig:Hybrid_MEMS_Equiv}(d), based on Ref. \cite{raghu2020spice}. The voltage across MEMS actuator $V_M$ is larger than the applied input voltage $V_{in}$ since the voltage across the ferroelectric $V_F$ is negative. This internal voltage amplification results in a lower pull-in voltage (corresponding to a lower energy barrier) in the hybrid MEMS actuator as compared to the standalone MEMS actuator.

\section{Hamiltonian of the Hybrid Actuator} \label{sec:formalism}

The Hamiltonian (total energy) $H_M$ of the standalone electrostatic MEMS actuator (Fig. \ref{fig:MEMS_1DOF}), is given by \cite{raghu2020charge}
\begin{equation} \label{eq:MEMS_EX}
H_{M}(x,\dot{x},t) = \underbrace{\frac{1}{2}~m~\dot{x}^2}_\text{Kinetic energy} + \underbrace{\frac{1}{2}~k~x^2 - \frac{1}{2}~\frac{\epsilon_o~A_M~V_M^2(t)}{(g_o - x)}}_\text{Potential energy}
\end{equation}
where $\dot{x}= \frac{dx}{dt}$ represents the velocity.

We assume the ferroelectric capacitor to behave as a single homogeneous domain. 
In the case where the ferroelectric material is inhomogeneous, the single domain assumption describes an averaged response, using an effective value of the ferroelectric coefficients \cite{li2015quantitative}. Indeed, literature (\cite{khan2015negative},\cite{saeidi2020nanowire}, \cite{khan2015negativeepi},  \cite{li2017negative,wang2017two,li2015quantitative,si2018steep}) reports the use of the single domain assumption to describe experimental results with different ferroelectrics for thicknesses upto $\sim$ 100 nm.  With the single domain assumption, the Landau-Khalatnikov (LK) equation \cite{masuduzzaman2014effective,salahuddin2008use,khan2015negative}  relates the voltage across the ferroelectric capacitor $V_F$ to the charge $q$ (where $q = \int i ~dt$) as 
\begin{equation} \label{eq:LK_Eqn}
V_F = -\alpha~q + \beta~q^3 + \gamma~q^5 
\end{equation} 
\begin{equation} \label{eq:Ferro_Coeff}
\alpha = -\frac{\alpha_F~t_F}{A_F}~, \beta = \frac{\beta_F~t_F}{A_F^3}~,\gamma=\frac{\gamma_F~t_F}{A_F^5}
\end{equation}
where $\alpha_F$, $\beta_F$ and $\gamma_F$ are ferroelectric anisotropy coefficients, $t_F$ and $A_F$ are the thickness and area of the ferroelectric respectively. The energy associated with the ferroelectric capacitor is given by 
\begin{equation} \label{eq:Ferro_UQ}
U_{F}(q) = -\frac{1}{2}~\alpha~q^2 + \frac{1}{4}~\beta~q^4 + \frac{1}{6}~\gamma~q^6 - V_F~q
\end{equation}


Note that the Eq. (\ref{eq:Ferro_UQ}) is written in terms of charge $q$, whereas Eq. (\ref{eq:MEMS_EX}) is described in terms of displacement $x$. In this work, we use displacement of the movable electrode as the common dynamical variable  to describe both the MEMS actuator and the ferroelectric capacitor. 
Both the ferroelectric capacitor and the MEMS actuator share the same charge $q$, since they are connected in series. Based on our earlier work on electrostatic MEMS actuators in Ref. \cite{raghu2020charge}, we relate the charge $q$ to the displacement $x$ of the movable electrode, using the mapping function 
\begin{equation} \label{eq:QX_Relation}
q = \frac{\epsilon_o~A_M~V_M(t)}{(g_o - x)}
\end{equation}

\noindent This mapping function is based on the charge-voltage relationship of a parallel plate capacitor. Using Eq. (\ref{eq:LK_Eqn}) and applying Kirchhoff's voltage law in Fig. \ref{fig:Hybrid_MEMS_Equiv}, Eq. (\ref{eq:QX_Relation}) can be rearranged to obtain 
\begin{equation}\label{eq:Q_solve}
\begin{split}
q^5 \left[\frac{\epsilon_o~A_M~\gamma}{g_o-x}\right] + q^3 \left[\frac{\epsilon_o~A_M~\beta}{g_o-x}\right]+q \left[1 - \frac{\epsilon_o~A_M~\alpha}{g_o-x}\right] \\ - \frac{\epsilon_o~A_M~V_{in}(t)}{g_o-x} = 0
\end{split}
\end{equation}

\begin{figure}[t]
    \centering
    \includegraphics[scale=1.0]{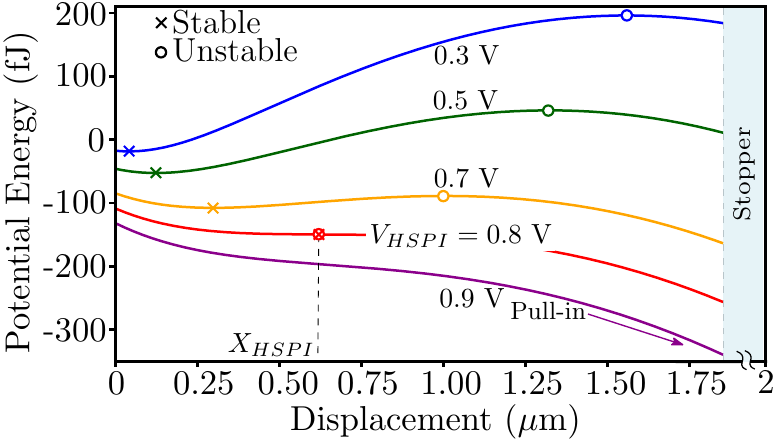}
    \caption{Potential energy-displacement plot depicting static pull-in. Stable and unstable equilibrium displacements coincide at static pull-in voltage, $V_{HSPI}=0.8$ V, with travel range $X_{HSPI}=0.62~\mu$m.}
    \label{fig:Hybrid_MEMS_Static}
\end{figure}

\begin{figure}[t]
    \centering
    \includegraphics[scale=1]{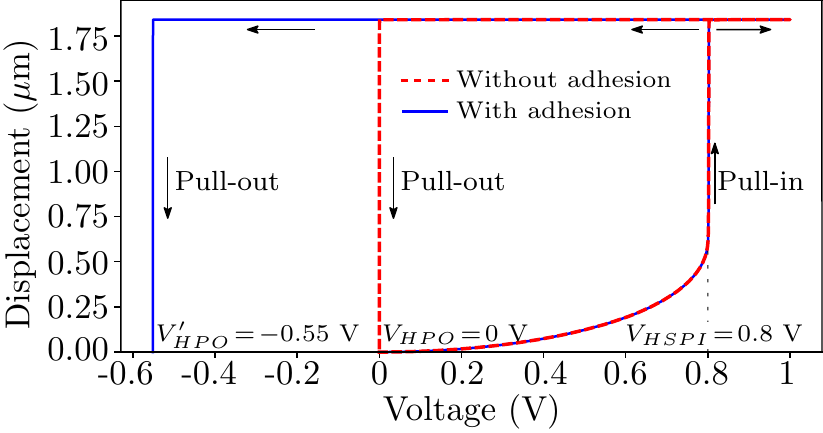}
    \caption{Numerical simulation of the static characteristics of the hybrid actuator  without (red dashed) and with (blue solid)  adhesion forces. Due to adhesion, the pull-out voltage is reduced from $V_{HPO}=0$ V to $V_{HPO}^\prime=-0.55$ V. However, the static pull-in voltage ($=0.8$ V) remains unchanged.}
    \label{fig:SPICE_adhesion}
\end{figure}

We  solve the above equation (discarding the complex roots), to obtain the charge $q$ as a function of the applied voltage and displacement. This charge is then substituted in Eq. (\ref{eq:LK_Eqn}) and Eq. (\ref{eq:Ferro_UQ}) to obtain the energy associated with the ferroelectric $U_F$, in terms of displacement.
We can thus write the Hamiltonian   of the hybrid actuator as
\begin{equation} \label{eq:Hybrid_MEMS_TotEnergy}
H_{H}(x,\dot{x},t) = U_{F}(x) + H_{M}(x,\dot{x},t)\\ 
\end{equation}
where both the ferroelectric and MEMS actuator are described in terms of the displacement of the MEMS actuator.  We will see later in Section \ref{sec:adhesion} that writing the Hamiltonian of the hybrid actuator in terms of $x$ allows us to include the effect of adhesion in a straightforward manner.

\begin{figure*}[t]
    \centering
    \includegraphics[scale=1]{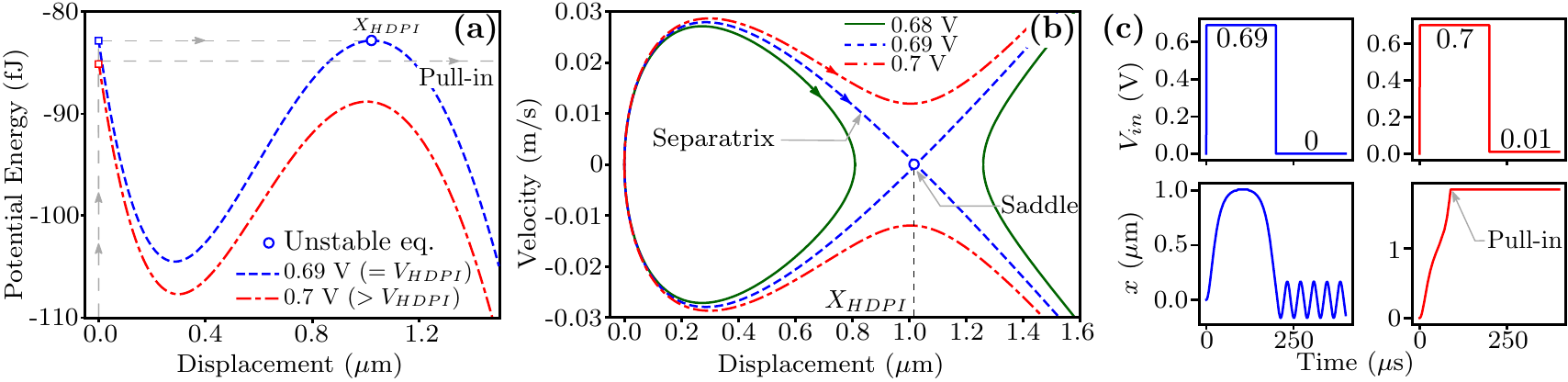}
    \caption{Analysis of dynamic pull-in using (a) Potential energy-displacement plot. Initial energy at $x=0$ equals the energy at the point of unstable equilibrium when the step input equals the dynamic pull-in voltage, $V_{HDPI}=0.69$ V. Any step input greater than $V_{HDPI}$ results in dynamic pull-in. (b) Phase-portrait. The trajectory becomes open when dynamic pull-in occurs. (c) Numerical simulation based on Ref. \cite{raghu2020spice}.}
    \label{fig:Hybrid_MEMS_Dynamic}
\end{figure*}

\section{Design of the Hybrid Actuator}\label{sec:design}

The effect of negative capacitance  has been reported in different ferroelectric materials like PbZr$_{0.2}$Ti$_{0.8}$O$_3$ (PZT) \cite{khan2015negative}, BiFeO$_3$ \cite{khan2015negativeepi}, Hafnium Zirconium Oxide (Hf$_{x}$Zr$_{1-x}$O$_2$) \cite{HoffmannNature2019, hoffmann2018demonstration,kobayashi2016experimental,mikolajick2021next, tasneem2021differential}, P(VDF-TrFE) \cite{ku2017transient,salvatore2012experimental} etc. In this work, we choose the parameters of  Hafnium Oxide (HfO$_2$) as a typical ferroelectric, with coercive field $E_c = 1$ MV/cm and remanant polarization $P_r = 9~\mu$C/cm$^2$ \cite{kobayashi2016device}. Note that  the essence of the analysis presented in this work does not depend on any particular choice of ferroelectric. The  ferroelectric capacitor must be designed so as that the combination of the ferroelectric and MEMS capacitors is stable at zero applied voltage \cite{masuduzzaman2014effective, raghu2020spice}. Note that the standalone MEMS actuator has  $V_{SPI} = 5.32$ V and $V_{PO}=4.88$ V (see Table \ref{tab:Table2}). We design the ferroelectric such that the hybrid actuator has a  static pull-in voltage of $0.8~$ V and pull-out voltage of $0$ V. 
Following Ref. \cite{masuduzzaman2014effective,raghu2020spice} the static pull-in voltage of the hybrid actuator $V_{HSPI}$ is given by 

\begin{subequations}
\label{eq:r_alpha} 
\begin{align}
V_{HSPI} &= r_{\alpha N}~ \sqrt{\frac{r_{\alpha N}}{r_{\beta N}} \cdot \frac{8~k~g_o^3}{27~\epsilon_0~A_M}} \text{ with}\\
r_{\alpha N} &= 1- {\frac{t_F~A_M~|\alpha_F|~\epsilon_o}{g_o~A_F}} \text{ and} \\
r_{\beta N} &= 1- \left[\frac{2~\beta_F~k~t_F~\epsilon_o^2 ~A_M^2}{A_F^3}\right]
\end{align} 
\end{subequations}
Pull-out of the hybrid actuator at $V_{HPO}=0$ V requires that the distance traveled by the movable electrode $(g_0 - h_s)= (r_{\alpha N}/r_{\beta N})~g_o$ \cite{masuduzzaman2014effective}. Using   the above equations, we obtain the required thickness $t_F$ and area $A_F$ of the ferroelectric, as listed in Table \ref{tab:Table1}.

\section{Analysis of the Hybrid Actuator} \label{sec:analysis}

From the Hamiltonian $H_H(x,\dot{x},t)$ in Eq. (\ref{eq:Hybrid_MEMS_TotEnergy}), we obtain the potential energy-displacement relation by setting $\dot{x}=0$. The hybrid actuator is analyzed using using the potential energy-displacement and phase-portrait (velocity-displacement) plots as explained in the following sections. 


\subsection{Static Pull-in}
The potential energy-displacement plot of the hybrid actuator shown in Fig. \ref{fig:Hybrid_MEMS_Static} explains static pull-in. For an applied voltage less than the static pull-in voltage, there are two equilibrium displacements: stable (denoted by $\times$) and unstable (denoted by $\circ$). These equilibrium displacements coincide when the input voltage equals the static pull-in voltage of the hybrid actuator, $V_{HSPI} = 0.8$ V. Correspondingly, the travel range of the hybrid actuator, $X_{HSPI} = 0.62~\mu$m. Beyond $V_{HSPI}$, the absence of any stable equilibrium displacement results in static pull-in, as depicted in Fig. \ref{fig:Hybrid_MEMS_Static}. The results obtained using the proposed framework exactly match with the analytical predictions (see Table \ref{tab:Table3}) and with the numerical simulations based on Ref. \cite{raghu2020spice} shown in Fig. \ref{fig:SPICE_adhesion}. The numerical simulations use a slowly varying ramp input (with $t_{inp}$ = 80 ms $\gg$ $t_{sys}$) as shown in Fig. \ref{fig:Hybrid_MEMS_Equiv}(d).


\subsection{Dynamic Pull-in}
Fig. \ref{fig:Hybrid_MEMS_Dynamic}(a) depicts dynamic pull-in in the hybrid actuator using potential energy-displacement plots. Note that the initial energy (energy at $x=0$) equals the energy at point of the unstable equilibrium for an applied step voltage of $0.69$ V. This corresponds to the dynamic pull-in voltage of the hybrid actuator, $V_{HDPI}$. Correspondingly, the dynamic pull-in displacement, $X_{HDPI}=1.01$ $~\mu$m, as depicted in Fig. \ref{fig:Hybrid_MEMS_Dynamic}(a). Any step input greater than $V_{HDPI}$ will result in dynamic pull-in because the initial energy is greater than the energy at the point of unstable equilibrium, as illustrated in Fig. \ref{fig:Hybrid_MEMS_Dynamic}(a). 

Dynamic pull-in can also be visualized using a phase-portrait (velocity-displacement plot). We plot the phase portrait by noting that total energy is conserved. Thus, for any step voltage $V_{in}$ applied at $t=0$ , the total energy $E_{total}$ can be obtained by setting  $x=0$, $\dot{x}=0$ (corresponding to the initial conditions) in Eq. (\ref{eq:Hybrid_MEMS_TotEnergy}). We then solve the implicit algebraic equation $H_H(x, \dot{x}) = E_{total}$ to obtain $\dot{x}$ for different values of $x$. The trajectory $\dot{x}(x)$ shows the  evolution of the system for a specific  applied step input, in the displacement-velocity phase plane. Note that time $t$ does not appear explicitly in Eq. (\ref{eq:Hybrid_MEMS_TotEnergy})  for a step input, provided $t > 0^+$.   The collection of trajectories for different applied voltages forms the phase-portrait.  

See Fig. \ref{fig:Hybrid_MEMS_Dynamic}(b). For step input less than $V_{HDPI}$, the trajectory is closed, indicating an oscillatory response. The phase-portrait shows a separatrix for a step input voltage of $V_{HDPI} = 0.69$ V. The separatrix runs through a saddle point that corresponds to the dynamic pull-in displacement, $X_{HDPI}=1.01~\mu$m. Any step input greater than $V_{HDPI}$ results in dynamic pull-in, which is characterized by the open trajectory in the phase-portrait. 

Results from numerical simulations  for dynamic  pull-in, based on Ref. \cite{raghu2020spice}, are shown in Fig. \ref{fig:Hybrid_MEMS_Dynamic}(c). They are obtained by applying a step voltage input (with $t_{inp} = 1$ ps $\ll$ $t_{sys}$).  The results from the graphical approach (Figs. \ref{fig:Hybrid_MEMS_Dynamic}(a), (b)) are in good agreement with the numerical simulations (summarized in Table \ref{tab:Table3}). Note that there are no analytical results for dynamic pull-in of the hybrid actuator.

\begin{figure}[t]
    \centering
    \includegraphics[scale=1]{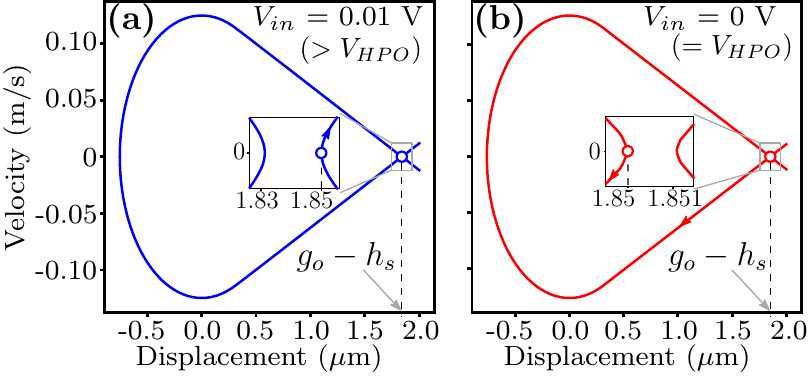}
    \caption{Analysis of pull-out using phase-portrait. Pull-out does not occur for input greater than the pull-out voltage, $V_{HPO}$, as displacement $g_o-h_s$ lies on the open trajectory. Pull-out occurs when input equals $V_{HPO}=0$ V as $g_o-h_s$ lies on the closed trajectory.}
    \label{fig:Hybrid_MEMS_Release}
\end{figure}
\subsection{Pull-out}
\begin{table}[b]
\caption{\label{tab:Table3}%
Summary of analysis of the hybrid actuator
}
\centering
\begin{adjustbox}{width=\columnwidth,center}
\begin{tabular}{lllll}
\toprule
\textrm{Parameter}&
\textrm{\makecell[l]{This \\ work}}&
\textrm{\makecell[l]{Numerical} }&
\textrm{\makecell[l]{Analytical}} \\
\midrule
\makecell[l]{Static pull-in voltage, $V_{HSPI}$} & $0.8$ V & $0.8$ V & $0.8$ V \\
\makecell[l]{Travel range, $X_{HSPI}$} & $0.62~\mu$m & $0.62~\mu$m & $0.62~\mu$m \\
\makecell[l]{Dynamic pull-in voltage, $V_{HDPI}$} & $0.69$ V & $0.69$ V & N.A.\\
\makecell[l]{Dynamic pull-in displacement, $X_{HDPI}$} & $1.01~\mu$m & $1.01~\mu$m & N.A.\\
\makecell[l]{Pull-out voltage, $V_{HPO}$} & $0$ V & $0$ V & $0$ V\\
\bottomrule
\end{tabular}
\end{adjustbox}
\end{table}
The pull-out of the hybrid actuator can  be visualized using the phase-portrait as shown in Fig. \ref{fig:Hybrid_MEMS_Release}(a), (b). Note that the actuator has already achieved pull-in, after travelling  a displacement of $x=g_o-h_s$, where $g_o$ is the air-gap and $h_s$ is the stopper height. Thus, for pull-out analysis, the initial displacement is at $x=g_o-h_s$, as opposed to the initial displacement of $x=0$, for the dynamic pull-in analysis. The hybrid actuator does not pull-out when the input is reduced upto $0.01$ V as the displacement $g_o-h_s$ lies on the open trajectory. However, when the input voltage is further reduced to $0$ V, the displacement $g_o-h_s$ lies on the closed trajectory, indicating pull-out. The closed trajectory represents sustained oscillations, after pull-out, in the absence of damping. Hence, the pull-out voltage of the hybrid actuator $V_{HPO}=0$ V. The estimated pull-out voltage of the hybrid actuator matches with the analytical prediction (see Table \ref{tab:Table3}) and with the numerical simulation shown in Fig. \ref{fig:SPICE_adhesion}. 




\section{Effect of Adhesion} \label{sec:adhesion}
\begin{figure*}[t]
    \centering
    \includegraphics[scale=1]{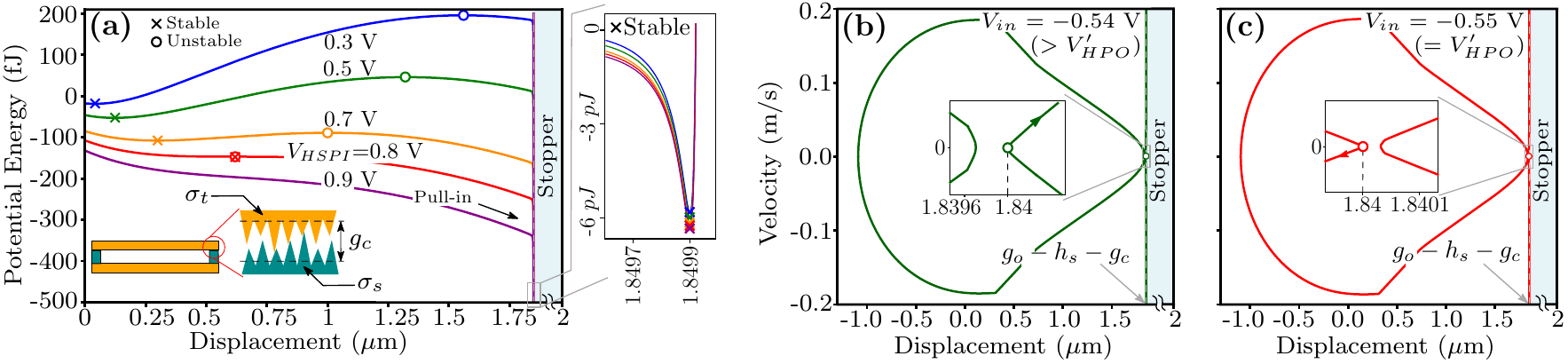}
    \caption{Effect of adhesion in the hybrid actuator. (a) Analysis of static pull-in using energy-displacement plot. The static pull-in voltage ($=0.8$ V) is not affected by adhesion. The plot near the contact is shown enlarged depicting the effect of adhesion. The inset in the main plot shows the effective gap at contact $g_c$ due to surface roughness ($\sigma_t$, $\sigma_s$ of the top electrode and stopper respectively), when the top electrode hits the stopper. (b), (c) Analysis of pull-out using phase-portrait. Note that the initial displacement for pull-out analysis is $g_o - h_s - g_c$ due to surface roughness. The pull-out voltage is reduced to $-0.55$ V as a result of adhesion.}
    \label{fig:Adhesion}
\end{figure*}

In this section, we analyze the pull-in and pull-out phenomena in the presence of adhesion (stiction force) between the contacting surfaces. When the beam is actuated so that the top electrode comes in contact with the stopper, adhesion  plays an important role. 
The proposed graphical approach based on energy landscape gives physical insight into understanding the effect of adhesion.  We model the adhesion between the contacting surfaces using the van der Waals force \cite{Delrio2005, granaldi2006dynamic, decuzzi2006bouncing,chakraborty2011experimental }, based on the Lennard-Jones potential and given by 
\begin{equation} \label{eq:LJforce}
F_{LJ}(x) = \frac{A_H A_C}{6 \pi}\left(\frac{1}{\delta^3(x)} - \frac{\Lambda^6}{\delta^9(x)}\right) 
\end{equation}
where $\delta(x) =  g_o - h_s - x$ is the gap between the top electrode and the stopper. $A_C$ is the area of contact.  $A_H$ is the Hamaker constant, and $\Lambda$ is the inter-atomic equilibrium distance. Eq. (\ref{eq:LJforce}) can also be expressed as 
\begin{equation} \label{eq:LJforce_2}
F_{LJ}(x) = \frac{C_1 A_C}{\delta^3(x)} - \frac{C_2 A_C}{\delta^9(x)}
\end{equation}
and $C_1, C_2$ are the attractive and repulsive constants, respectively, with typical values  $C_1 = 10^{-20}$ Nm, $C_2 = 10^{-80}$ Nm$^7$  \cite{granaldi2006dynamic}. The corresponding Lennard-Jones potential is given by 
\begin{equation}
U_{LJ}(x) = \frac{-~C_1 A_C}{2~(g_o - h_s - x)^2} + \frac{~C_2 A_C}{8~(g_o - h_s - x)^8}
\end{equation} 
Since $U_{LJ}$ is a function only of $x$, we can add it directly to the Hamiltonian of the hybrid actuator $H_{H}(x,\dot{x},t)$. The Hamiltonian of the hybrid MEMS actuator is now modified to include the Lennard-Jones potential as 
\begin{equation} \label{eq:adh}
H_{HLJ}(x,\dot{x},t) = H_{H}(x,\dot{x},t) + U_{LJ}(x)
\end{equation}

The effect of adhesion on the pull-in voltage is analyzed using the potential energy-displacement plot using Eq. (\ref{eq:adh}) with $\dot{x}=0$. As depicted in Fig. \ref{fig:Adhesion}(a), the static pull-in voltage ($=0.8$ V) is not changed due to adhesion. This is because  van der Waals force is a short-range force which does not change the energy landscape in regions away from the contact. 

The numerical simulation of the static characteristics in Fig. \ref{fig:SPICE_adhesion}, based on Ref. \cite{raghu2020spice}, also confirms that the static pull-in voltage is unaffected due to adhesion. We have modified the simulation framework in Ref. \cite{raghu2020spice} to include the effect of adhesion. Adhesion is implemented as an additional sub-circuit which models the van der Waals force (Eq. (\ref{eq:LJforce_2})). The numerical model of the standalone MEMS actuator now estimates the acceleration $a$ based on the following force-balance equation
\begin{equation}
a = \frac{F_{elec} + F_{LJ} - F_{mech}}{m}
\end{equation}
where $F_{elec},F_{LJ}$ and $F_{mech}$ are the electrostatic, van der Waals and the mechanical restoring forces, respectively. The estimated acceleration is integrated to compute the velocity $\dot{x}$, which is again integrated to obtain the displacement $x$. The integration is performed by a built-in function available in the circuit simulator. The estimated velocity and displacement are given in a feedback loop  to obtain the stable solution of the electrode displacement for an applied input voltage.

Note that a deep energy well, with a stable minima very close to the stopper location ($\approx g_o - h_s$) is created  due to adhesion, as shown in Fig. \ref{fig:Adhesion}(a). When the applied voltage exceeds the pull-in voltage, pull-in occurs and the beam comes in contact with the stopper. Due to the roughness of the contact surface, there are small asperities distributed all over the contact area [see inset in Fig. \ref{fig:Adhesion}(a)]. As a result, there exists an effective gap at contact, $g_c$, between the two contacting surfaces. Assuming $\sigma_t$ and $\sigma_s$ are the standard deviations of the thickness of the top electrode and the height of the stopper respectively, we can define $g_c = \sqrt{\sigma_t^2 + \sigma_s^2}$\cite{chakraborty2011experimental,hariri2006modeling}.  Therefore, after pull-in, owing to the surface roughness, the top electrode settles effectively at $g_o - h_s - g_c$. The effective gap is a random variable that varies across different fabrication runs. For the chosen dimensions of the MEMS beam, we assume $g_c=10$ nm, based on Refs. \cite{shekhar2017surface,chakraborty2011experimental}. 

The pull-out voltage is analyzed using the phase-portrait, as shown in Fig. \ref{fig:Adhesion}(b), (c). As explained above, after pull-in, the top electrode settles at $g_o-h_s-g_c$. Therefore, for the analysis of pull-out using the phase portrait, the initial displacement to determine the total energy is $g_o-h_s-g_c$. Contrast this with the case without adhesion and surface roughness, wherein the initial displacement for pull-out analysis is $g_o-h_s$ [see Fig. \ref{fig:Hybrid_MEMS_Release}(a), (b)]. As shown in Fig. \ref{fig:Adhesion}(b), (c), the displacement $g_o - h_s - g_c$ is on the closed trajectory when the input is $-0.55$ V. Thus, adhesion reduces the pull-out voltage from $V_{HPO}=0$ V (without adhesion; see Table \ref{tab:Table3}) to $V_{HPO}^\prime=-0.55$ V. We also confirm this reduction in the pull-out voltage using the numerical simulation of the static characteristics, as shown in Fig. \ref{fig:SPICE_adhesion}. Thus, the pull-in voltage is unaffected and the pull-out voltage is reduced due to adhesion.
\begin{figure}[t]
    \centering
    \includegraphics[scale=1]{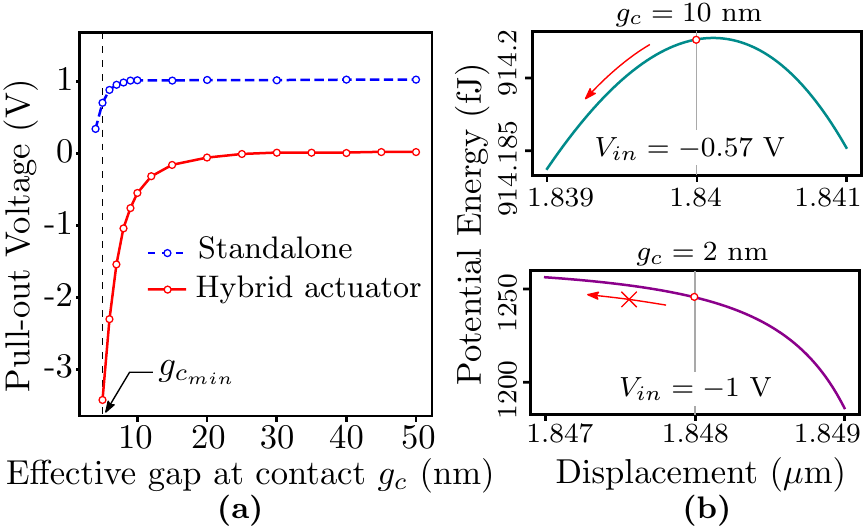}
    \caption{(a) Variation in the pull-out voltage with respect to the change in the effective gap at contact $g_c$. As effective gap increases, the pull-out voltage tends towards the electrostatically estimated (without adhesion) value. (b) Energy-displacement plots for $g_c=10$ nm and 2 nm. The presence of an energy barrier prevents pull-out for $g_c=2$ nm ($< g_{c_{min}}$). Note that the \textcolor{red}{$\circ$} represents $(g_o - h_s - g_c)$ in both cases.} 
    \label{fig:eff_gap}
\end{figure}
\begin{figure}[b]
    \centering
    \includegraphics[scale=1]{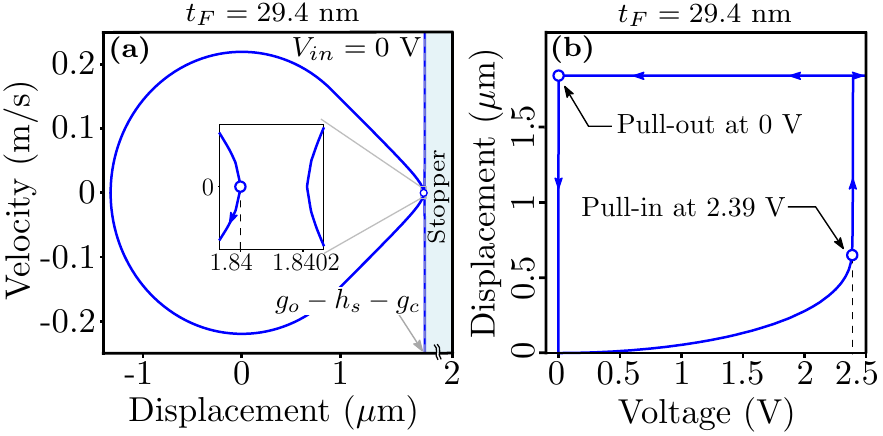}
    \caption{(a) Phase-portrait depicting that the pull-out voltage, in the presence of adhesion, is $0$ V, achieved with a reduced ferroelectric thickness of $t_F=29.4$ nm. (b) Numerical simulation of the static characteristics of the hybrid actuator with $t_F=29.4$ nm.}
    \label{fig:ferro_thickness}
\end{figure}

We now study the variation in the pull-out voltage with respect to the change in the effective gap at contact ($g_c$), as shown in Fig. \ref{fig:eff_gap}(a). The trend observed in the hybrid actuator is similar to the effect of adhesion on the pull-out voltage in the standalone actuator (for example, see Ref. \cite{chakraborty2011experimental}). As shown in Fig. \ref{fig:eff_gap}(a), the pull-out voltage tends towards the electrostatically estimated value (without adhesion) with an increase in  $g_c$. This is because adhesion force becomes negligible for higher values of $g_c$. As in the case of the standalone actuator, there exists a minimum effective gap at contact ($g_{c_{min}}$), in the hybrid actuator, below which pull-out does not occur in the presences of adhesion. For example, as shown in Fig. \ref{fig:eff_gap}(b), pull-out occurs for $g_c=10$ nm ($> g_{c_{min}}$) with $V_{in}=-0.57$V ($< V_{HPO}^\prime$). However, the presence of an energy barrier at $g_o-h_s-g_c = 1.848~\mu$m for $g_c=2$ nm ($< g_{c_{min}}$) prevents pull-out. Thus, the effective gap at contact $g_c$ and hence the surface roughness plays a significant role in determining the pull-out behaviour. For instance, pull-out can be facilitated by increasing the surface roughness \cite{tas1996stiction} thereby reducing stiction. 

We predict that the pull-out voltage can be brought back to $0$ V, even in the presence of adhesion, by tailoring the ferroelectric thickness $t_F$. It has been predicted that both pull-in and pull-out voltages increase with reduction in the ferroelectric thickness \cite{raghu2020spice,choe2018ferroelectric}. By looking at the slope of the potential energy-displacement plot at $g_o - h_s - g_c$, we predict that a reduction of the ferroelectric thickness to $29.4$ nm eliminates the barrier for pull-out at zero applied voltage, as shown in Fig. \ref{fig:ferro_thickness}(a), (b). This is, however, accompanied with an increased pull-in voltage of $2.39$ V. Nevertheless this increased pull-in voltage is  is still lower than the pull-in voltage of the standalone MEMS actuator (=$5.32$ V).   

\section{Conclusion} \label{sec:conclusion}
To summarize, we have proposed a physics-based framework based on the energy-landscape to systematically analyze the static pull-in, dynamic pull-in and pull-out phenomena of the ferroelectric negative capacitance-hybrid MEMS actuator. Based on the proposed framework, we illustrate the low-voltage operation of the hybrid actuator for static and step inputs. The results obtained are in good agreement with  analytical predictions and numerical simulations. We also include the effect of adhesion in the framework.  We show that the pull-in voltage is not affected, while the pull-out voltage is reduced due to adhesion. Since the proposed framework employs graphical energy-displacement and phase-portrait plots, it serves as an easy and quick design and analysis tool. The scope of the proposed framework can further be extended to include other effects that are defined in terms of displacement (for example, fringing capacitance of the MEMS actuator \cite{nemirovsky2001methodology}).  This work should aid in the study of ferroelectric negative capacitance for electrostatic MEMS applications. 

\section*{acknowledgments}
AA acknowledges support from SERB (Science and
Engineering Research Board, Government of India) through
SRG/2019/001229.

\bibliographystyle{IEEEtran}       
\bibliography{TED_Hybrid_Energy_Adhesion}

\end{document}